\documentclass[12pt,english,floatfix,showkeys,superscriptaddress,aps,prd,preprint]{revtex4}
\usepackage[latin1]{inputenc}
\usepackage[T1]{fontenc}
\usepackage{lmodern}
\usepackage{amsmath}
\usepackage{amssymb}
\usepackage{graphicx}
\usepackage{float}
\usepackage{esint}
\usepackage{longtable}
\usepackage{dcolumn}
\usepackage{babel}
\usepackage{csquotes}
\usepackage{color}
\usepackage{slashed}
\usepackage{simplewick}
\usepackage{amsmath,latexsym}

\usepackage{cancel}

\usepackage{hyperref}
\hypersetup{
    colorlinks,
    citecolor=blue,
    filecolor=green,
    linkcolor=purple,
    urlcolor=red,
}

\usepackage{slashed}

\usepackage{hyperref}
\hypersetup{colorlinks,breaklinks,
			citecolor=[rgb]{0,0.0,1.0},
            urlcolor=[rgb]{0.0,0.0,1.0},
            linkcolor=[rgb]{0,0.5,0.9}}

\begin{document}

\title{Vacuum polarization and induced Maxwell and Kalb-Ramond effective action in very special relativity}

%%%%%%%%%%%%%%%%%%%%%%
\author{Roberto V. Maluf}
\email{r.v.maluf@fisica.ufc.br}
\affiliation{Universidade Federal do Cear\'a (UFC), Departamento de F\'isica,\\ Campus do Pici, Fortaleza - CE, C.P. 6030, 60455-760 - Brazil.}
\affiliation{Departamento de F\'{i}sica Te\'{o}rica and IFIC, Centro Mixto Universidad de Valencia - CSIC. Universidad
de Valencia, Burjassot-46100, Valencia, Spain.}

%%%%%%%%%%%%%%%%%%%%%%%%%%%%%%%%%%%%%%%%%%%%%%%%%%%%%%%%%%%%%%%%%%%%%%
\author{Gonzalo J. Olmo}
\email{gonzalo.olmo@uv.es}
\affiliation{Departamento de F\'{i}sica Te\'{o}rica and IFIC, Centro Mixto Universidad de Valencia - CSIC. Universidad
de Valencia, Burjassot-46100, Valencia, Spain.}
\affiliation{Universidade Federal do Cear\'a (UFC), Departamento de F\'isica,\\ Campus do Pici, Fortaleza - CE, C.P. 6030, 60455-760 - Brazil.}

%%%%%%%%%%%%%%%%%%%%%%%%%%%%%%%%%%%%%%%%%%%%%%%%%%%%%%%%%%%%%%%%%%%%%%

\date{\today}

\begin{abstract}
This work investigates the implications of very special relativity (VSR) on the calculation of vacuum polarization for fermions in the presence of Maxwell and Kalb-Ramond gauge fields in four-dimensional spacetime. We derive the $SIM(2)$-covariant gauge theory associated with an Abelian antisymmetric 2-tensor and its corresponding field strength. We demonstrate that the free VSR-Kalb-Ramond electrodynamics is equivalent to a massive scalar field with a single polarization. Furthermore,  we determine an explicit expression for the effective action involving Maxwell and Kalb-Ramond fields due to fermionic vacuum polarization at one-loop order. The quantum corrections generate divergences free of nonlocal terms only in the VSR-Maxwell sector. At the same time, we observe UV/IR mixing divergences due to the entanglement of VSR-nonlocal effects with quantum higher-derivative terms for the Kalb-Ramond field. However, in the lower energy limit, the effective action can be renormalized like in the Lorentz invariant case. 
\end{abstract}

%\pacs{04.70.-s,04.50.Kd,11.30.Cp,04.60.-m}

\maketitle

%%%%%%%%%%%%%%%%%%%%%%%%%%%%%%%%%%%%%%%%%%%%%%%%%%%%%%%%%%%%%%%%%%%%%%

\section{Introduction}

In 2006, Cohen and Glashow proposed a modification to the theory of special relativity that preserves the familiar energy-momentum dispersion relation while breaking the invariance under the complete Lorentz group, $SO(1,3)$ \cite{Cohen:2006ky}. They recognized that specific subgroups of the Lorentz group can still produce conservation laws and reproduce the well-known effects of special relativity (SR). Among these subgroups, the $HOM(2)$ and $SIM(2)$ groups meet these requirements. The former subgroup, known as the Homothety group, consists of the boost generator $K_{z}$ and the generators $T_{1} = K_{x}+ J_{y}$, $T_{2}=K_{y}-J_{x}$. These generators together form a group that is isomorphic to the group of translations in the plane. The second subgroup, known as the similitude group $SIM(2)$, is an enhanced version of the $HOM(2)$ group obtained by including the generator $J_{z}$. These subgroups do not admit invariant tensors that can act as constant background tensor fields, as seen in other theories of Lorentz violations \cite{Colladay:1998fq}. Hence, the breakdown of $SO(3,1)$ into either $HOM(2)$ or $SIM(2)$ cannot be explained through local symmetry-breaking operators. This intriguing theory of relativity came to be known as ``very special relativity'' (VSR) \cite{Fan:2006nd}.

A key attribute of VSR is that the $SIM(2)$ generators preserve both the speed of light and the null 4-vector $n_{\mu}=(1, 0, 0, 1)$, thereby establishing a preferred direction in space. However, VSR algebras do not support discrete symmetry operators such as $P$, $T$, $CP$, and $CT$. Including any of these operators would result in the full Lorentz algebra \cite{Lee:2015tcc}. In this way, the lack of discrete symmetries may result in violations of unitarity and causality in quantum field theories. To circumvent this issue, Cohen and Glashow proposed the inclusion of non-local operators containing ratios of contractions of $n_{\mu}$ in order to construct a unitary $SIM(2)$-invariant field theory. The non-local nature of VSR gives rise to a remarkable phenomenon of mass generation. Indeed, when a non-local operator like $n_{\mu}/(n\cdot \partial)$ is added in the momentum operator for a massless fermion, it results in a corresponding Klein-Gordon equation with a mass term proportional to the non-local coupling constant. This impressive property has led to proposals suggesting a VSR explanation for neutrino mass \cite{Cohen:2006ir} and dark matter \cite{Ahluwalia:2010zn}.

Since its formulation, a large number of investigations in VSR theories have been developed in recent years, addressing distinct physical scenarios: fermion systems \cite{Bernardini:2007uj,Maluf:2014yca,Koch:2022jcd}, Maxwell \cite{Cheon:2009zx,Alfaro:2013uva}, Maxwell-Chern-Simons \cite{Bufalo:2016lfq} and axion \cite{Bufalo:2017yxs} electrodynamics, Lorentz violation \cite{Alfaro:2013uga}, curved spacetime \cite{Muck:2008bd}, noncommutativity \cite{Sheikh-Jabbari:2008ybm, Das:2010cn},  linearized gravity \cite{Alfaro:2022vpt}, Finsler geometry \cite{Gibbons:2007iu, Kouretsis:2008ha}, supersymmetry \cite{Cohen:2006sc}, radiative corrections \cite{Alfaro:2019xzh,Alfaro:2019koq,Bufalo:2020cst,Haghgouyan:2022dvq}. These several studies have served both to elucidate the effects engendered by VSR non-local terms and to set up upper bounds on the VSR mass coefficients.

In particular, the VSR contributions to the induced effective action in the context of Maxwell-Chern-Simons electrodynamics have been studied in previous works \cite{Bufalo:2020cst,Haghgouyan:2022dvq}. In these studies, the authors employed the Mandelstam-Leibbrandt prescription \cite{Leibbrandt:1983pj, Leibbrandt:1987qv, LeibbrandtBook}, adapted to the VSR case by Alfaro \cite{Alfaro:2016pjw, Alfaro:2017umk}, to handle the UV/IR mixing divergences that arise in the loop integrals. Another notable gauge field theory in four-dimensional Minkowski spacetime is built from an antisymmetric 2-tensor known as the Kalb-Ramond field \cite{Kalb:1974yc}. The Kalb-Ramond field is relevant in several contexts and has been extensively explored in the literature, including string theories \cite{Rey:1989xj}, quantum field theory \cite{Deguchi:1998xp, Lahiri:1999uc, Mukhopadhyay:2014dva},  supersymmetry \cite{Almeida:2015yna,Gama:2022nue}, Lorentz symmetry violation \cite{Altschul:2009ae, Hernaski:2016dyk, Maluf:2018jwc}, dualities \cite{Menezes:2002nj, Maluf:2020fch}, black hole and wormhole solutions \cite{Lessa:2019bgi, Maluf:2021ywn}, cosmology \cite{Maluf:2021eyu}, and brane world scenarios \cite{Cruz:2013zka, Belchior:2023gmr, Moreira:2023pes}. 

The study of the classical action for antisymmetric tensor fields in the context of VSR was initially addressed in Ref. \cite{Upadhyay:2015yyt}. However, a comprehensive analysis of the role of VSR in the quantum corrections of this field is still lacking in the literature. In this work, we propose a systematic procedure to construct an antisymmetric 2-tensor gauge field that incorporates VSR non-local operators and use it to derive the associated $SIM(2)$-invariant classical action. Subsequently, we obtain the free equation of motion in the VSR-Kalb-Ramond electrodynamics, and explicitly determine the resulting degrees of freedom. Furthermore, we explore, for the first time, the induced corrections to the effective action of the Maxwell and Kalb-Ramond gauge fields in the context of VSR. We obtain the effective Lagrangian density at the one-loop order by integrating the fermionic fields and calculating the vacuum polarization Feynman diagrams. We demonstrate that the divergent terms can be renormalized in the low-energy limit by appropriately rescaling the fields, masses, and coupling parameters in the model. Moreover, our results recover those obtained in the literature for the Lorentz invariant limit \cite{HariDass:2001dp}.

The present work is organized as follows. In Sec. \ref{sec2}, we review the main aspects of VSR applied to the vector gauge field. In Sec. \ref{sec3}, we propose a procedure to derive the $SIM(2)$-invariant Lagrangian density for the Kalb-Ramond field and analyze its free physical modes. In Sec. \ref{sec4}, we calculate the induced effective actions of the Maxwell and Kalb-Ramond fields, evaluate the Feynman diagrams for the two-point gauge functions, and examine the general tensorial form of the finite and divergent induced terms as well as the renormalization issues. Our final comments are presented in Section \ref{sec5}.

%%%%%%%%%%%%%%%%%%%%%%%%%%%%%%%%%%%%%%%%%%%%%%%%%%%%%%%%%%%%%%%%%%%%%%%%%%
\section{SIM(2)-invariant Maxwell gauge theory \label{sec2}}
To establish a consistent framework for our upcoming calculations, we will review the $SIM(2)$-invariant gauge vector theory described in Refs. \cite{Cheon:2009zx,Alfaro:2013uva,Alfaro:2019koq}. The VSR-modified Maxwell electrodynamics is a $U(1)$ gauge theory that involves a $1$-form gauge potential $A_{\mu}(x)$ and a matter field $\psi(x)$ that acts as the source of $A_{\mu}(x)$. This theory obeys the following gauge transformations:
\begin{eqnarray}
A_{\mu}(x)&\rightarrow A_{\mu}(x)+\tilde{\partial}_{\mu}\Lambda(x),\label{SIM2gaugeA}\\
\psi(x)&\rightarrow\exp\left\{ ie\Lambda(x)\right\} \psi(x),
\end{eqnarray}where $\Lambda(x)$ is an arbitrary $0$-form field.  The wiggle derivative operator is defined by
\begin{equation} \tilde{\partial}_{\mu}=\partial_{\mu}+\frac{1}{2}\frac{m_{A}^{2}}{n\cdot\partial}n_{\mu}, \end{equation}
where $m_{A}$ is a constant parameter with mass dimension, and $n^{\mu}=(1,0,0,1)$ is a fixed null vector present in VSR theories and select a preferred direction.

The covariant derivative for VSR-Maxwell electrodynamics is given by
\begin{equation}
\mathcal{D}_{\mu}=\partial_{\mu}-ie\left(A_{\mu}-\frac{1}{2}m_{A}^{2}n_{\mu}\left(\frac{1}{(n\cdot\partial)^{2}}n\cdot A\right)\right)
\end{equation}and it is constructed by demanding the fundamental property of transforming as $\psi$ does under infinitesimal gauge transformations \cite{Alfaro:2013uva}:
\begin{equation}
\delta(\mathcal{D}_{\mu}\psi)=ie\Lambda\mathcal{D}_{\mu}\psi.
\end{equation}
We can compute the field strength related to $\mathcal{D}_{\mu}$ as follows:
\begin{equation}
[\mathcal{D}_{\mu},\mathcal{D}_{\nu}]\psi=-ie\mathcal{F}_{\mu\nu}\psi.
\end{equation}This gives us the expression for the field strength tensor:
%\begin{equation}
%\mathcal{F}_{\mu\nu}=\partial_{\mu}A_{\nu}-\partial_{\nu}A_{\mu}+\frac{1}%{2}m_{A}^{2}n_{\mu}\left(\frac{1}{(n\cdot\partial)^{2}}\partial_{\nu}(n\cdot A)\right)-\frac{1}{2}m_{A}^{2}n_{\nu}\left(\frac{1}{(n\cdot\partial)^{2}}\partial_{\mu}(n\cdot A)\right).\label{Ftilde}
%\end{equation}
\begin{equation}
\mathcal{F}_{\mu\nu}=\partial_{\mu}A_{\nu}-\partial_{\nu}A_{\mu}+\frac{1}{2}m_{A}^{2}N_{\mu}\partial_{\nu}(N\cdot A)-\frac{1}{2}m_{A}^{2}N_{\nu}\partial_{\mu}(N\cdot A).\label{Ftilde}
\end{equation}where we introduce the notation $N_{\mu}\equiv n_{\mu}/n\cdot \partial$ for the nonlocal vector operator.

It is noteworthy that $\mathcal{F}_{\mu\nu}$ is invariant under the gauge transformation 
\begin{equation}
\tilde{A}_{\mu}\to\tilde{A}_{\mu}+\partial_{\mu}\Lambda,\label{WiggleGaugeA}
\end{equation}where we define the wiggle gauge vector $\tilde{A}_{\mu}$ by
%\begin{equation}
%\tilde{A}_{\mu}=A_{\mu}-\frac{1}{2}m_{A}^{2}n_{\mu}\left(\frac{1}{(n\cdot\partial)^{2}}n\cdot A\right),\label{Atilde}
%\end{equation}
\begin{equation}
    \tilde{A}_{\mu}=A_{\mu}-\frac{1}{2}m_{A}^{2}N_{\mu}\left(N\cdot A\right),\label{Atilde}
\end{equation}such that $\mathcal{F}_{\mu\nu}=\partial_{\mu}\tilde{A}_{\nu}-\partial_{\nu}\tilde{A}_{\mu}$.

We observe that by applying a field redefinition $A_{\mu}\rightarrow A_{\mu}+\frac{1}{2}m_{A}^{2}N_{\mu}\left(N\cdot A\right)$, one can eliminate the modification by the VSR-nonlocal terms. After this redefinition, we obtain the same field strength tensor $\mathcal{F}_{\mu\nu}\rightarrow F_{\mu\nu}=\partial_{\mu}A_{\nu}-\partial_{\nu}A_{\mu}$ and the same covariant derivative gauge transformation as in standard electrodynamics, given by
\begin{equation}
\mathcal{D}_{\mu}\psi\rightarrow D_{\mu}\psi=\partial_{\mu}\psi-ieA_{\mu}\psi.
\end{equation}

On the other hand, we can define a new $SIM(2)$-gauge invariant field strength $\tilde{F}_{\mu\nu}$, which is constructed using the wiggle derivative, namely, 
\begin{equation}
\tilde{F}_{\mu\nu}=\tilde{\partial}_{\mu}A_{\nu}-\tilde{\partial}_{\nu}A_{\mu},\label{Ftilda1}
\end{equation}
and it can be expressed explicitly as
\begin{equation}
\tilde{F}_{\mu\nu}=\mathcal{F}_{\mu\nu}+\frac{1}{2}m_{A}^{2}\left[n_{\mu}\frac{1}{(n\cdot\partial)^{2}}(n^{\alpha}\mathcal{F}_{\alpha\nu})-n_{\nu}\frac{1}{(n\cdot\partial)^{2}}(n^{\alpha}\mathcal{F}_{\alpha\mu})\right],\label{Ftilda2}
\end{equation}
which shows that $\tilde{F}_{\mu\nu}$ is also invariant under 
the gauge transformation (\ref{WiggleGaugeA}).

The present analysis allows us to construct a $SIM(2)$-invariant Lagrangian density for the field $A_{\mu}$ that is also invariant under standard gauge transformations. As pointed out in Ref. \cite{Alfaro:2019koq}, this is an important result because this Lagrangian generates a mass term for the field $A_{\mu}$ without breaking the original gauge symmetry of the theory.

According to the definition (\ref{Ftilda1}), $\tilde{F}_{\mu\nu}$ is not Lorentz invariant but instead $SIM(2)$ invariant. Moreover, as shown in result (\ref{Ftilda2}), it is also invariant under transformation (\ref{WiggleGaugeA}). So, we can construct a VSR gauge-invariant Lagrangian density as follows:
\begin{equation}
\mathcal{L}_{gauge}=-\frac{1}{4}\tilde{F}_{\mu\nu}\tilde{F}^{\mu\nu}.
\end{equation}
Therefore, from Eq. (\ref{Ftilda2}), this Lagrangian takes the form
\begin{equation}
\mathcal{L}_{gauge}=-\frac{1}{4}\mathcal{F}_{\mu\nu}\mathcal{F}^{\mu\nu}+\frac{1}{2}m_{A}^{2}n_{\mu}\left(\frac{1}{n\cdot\partial}\mathcal{F}^{\mu\nu}\right)n^{\alpha}\left(\frac{1}{n\cdot\partial}\mathcal{F}_{\alpha\nu}\right).
\end{equation}
Finally, by applying a field redefinition from $A_{\mu}\rightarrow A_{\mu}+\frac{1}{2}m_{A}^{2}n_{\mu}((n\cdot\partial)^{-2}(n\cdot A))$, so that $\mathcal{F}_{\mu\nu}\rightarrow F_{\mu\nu}=\partial_{\mu}A_{\nu}-\partial_{\nu}A_{\mu}$, we obtain the desired result
\begin{equation}
\mathcal{L}_{gauge}=-\frac{1}{4}F_{\mu\nu}F^{\mu\nu}+\frac{1}{2}m_{A}^{2}n_{\mu}\left(\frac{1}{n\cdot\partial}F^{\mu\nu}\right)n^{\alpha}\left(\frac{1}{n\cdot\partial}F_{\alpha\nu}\right).\label{Lgauge}
\end{equation}It is interesting to notice that if we start from $-1/4\mathcal{F}_{\mu\nu}\mathcal{F}^{\mu\nu}$ to define our Lagrangian, the above field redefinition will withdraw the VSR effects. Furthermore, as shown in Ref. \cite{Alfaro:2019koq}, we can apply the Lorentz gauge $\partial_{\mu}A^{\mu}=0$ plus the subsidiary gauge condition $N\cdot A=0$ into  
the equation of motion obtained from (\ref{Lgauge}) and we find 
\begin{equation}
(\partial^{2}+m_{A}^{2})A^{\nu}=0.
\end{equation}Hence, in the VSR scenario, we obtain a massive gauge field with two physical degrees of freedom, which is in contrast to the Proca case where the mass term $m_{A}^2A^{\mu}A_{\mu}$ is not gauge invariant and has three degrees of freedom.

%%%%%%%%%%%%%%%%%%%%%%%%%%%%%%%%%%%%%%%%%%%%%%%%%%%%%%%%%%%%%%%%%%%%%%%%%%%%%%%%%%%%%%%%
\section{Kalb-Ramond electrodynamics in VSR \label{sec3}}

In this section, we investigate the issue of constructing a $SIM(2)$-invariant action for the Kalb-Ramond field. As we will see, this is possible even when the Kalb-Ramond field does not carry matter charge, i.e., when it is not minimally coupled to matter fields and does not possess any associated covariant derivatives.
\subsection{Setup}

Let us start by defining the Lagrangian density that describes the dynamics for an antisymmetric 2-tensor $B_{\mu\nu}$  in $4D$ Minkowski spacetime,
\begin{equation}
\mathcal{L} = -\frac{1}{12} H_{\mu\nu\alpha}H^{\mu\nu\alpha} - \frac{1}{2}B_{\mu\nu}J^{\mu\nu},\label{LKR}
\end{equation} where
\begin{equation} 
H_{\mu\nu\alpha}=\partial_{\mu}B_{\nu\alpha} + \partial_{\alpha}B_{\mu\nu} + \partial_{\nu}B_{\alpha\mu},\label{Hfield}
\end{equation}
is the field strength tensor associated with $B_{\mu\nu}$, and $J^{\mu\nu}$ is an antisymmetric conserved current due to the coupling to the matter  \cite{Altschul:2009ae}.  
The field strength $H_{\mu\nu\alpha}$ corresponds to the components of an exact 3-form field $H$, which is constructed using the exterior derivative from the 2-form $B$ associated with $B_{\mu\nu}$. This field strength satisfies the identity
\begin{equation}
\partial_{\kappa}H_{\lambda\mu\nu} - \partial_{\lambda}H_{\mu\nu\kappa} + \partial_{\mu}H_{\nu\kappa\lambda} - \partial_{\nu}H_{\kappa\lambda\mu} = 0,
\label{Hid}
\end{equation}which follows from the fact that an exact 3-form is closed \cite{Nakahara:2003nw}.

The Lagrangian (\ref{LKR}) is the simplest which can be constructed by demanding parity-even and invariance under the $U(1)$ gauge transformation:
\begin{equation}
B_{\mu\nu}(x) \rightarrow B_{\mu\nu}(x) + \partial_{\mu} \Sigma_{\nu}(x) - \partial_{\nu} \Sigma_{\mu}(x), \label{gaugeKR}
\end{equation}where $\Sigma_{\mu}$ is an arbitrary vector field.  The field $\Sigma_{\mu}$ also exhibits an extra gauge invariance given by 
\begin{equation}
\Sigma_{\mu}(x) \rightarrow \Sigma_{\mu}(x) + \partial_{\mu}\phi(x),\label{gaugeSigma}
\end{equation} with $\phi$ being an arbitrary scalar field. This latter transformation leaves Eq. (\ref{gaugeKR}) unchanged.

In general, the current $J_{\mu\nu}$ is constructed from other dynamical fields which involve extended objects of the type found in the string theory \cite{Kalb:1974yc}. For the sake of simplicity, we will not consider a string matter field for the source of $B_{\mu\nu}(x)$ in this work. In what follows, our attention will be focused only on the kinetic part of the Lagrangian density (\ref{LKR}), such that the matter coupling, represented by $J_{\mu\nu}$, will be turned off.

%%%%%%%%%%%%%%%%%%%%%%%%%%%%%%%%%%%%%%%%%%%%%%%%%%%%%%%%%%%%%%%%%%%%%%%%%%%%%%%%%%%%%%
\subsection{$SIM(2)$-covariant Kalb-Ramond gauge theory}

For the $SIM(2)$-invariant generalization of the Kalb-Ramond Lagrangian (\ref{LKR}), we expect that the gauge symmetry (\ref{gaugeKR}) modified by the nonlocal vector operator $N_{\mu}\equiv n_{\mu}/n\cdot \partial$ will play a crucial role. As we saw in the Maxwell case, both $\mathcal{F}_{\mu\nu}$ and $\tilde{F}_{\mu\nu}$ are invariant under the standard gauge transformation, and the connection between the two kinds of field strengths is made through the vector potential $\tilde{A}_{\mu}$.

Hence, motivated by our earlier analysis, we will construct a $\tilde{B}_{\mu\nu}$ field that satisfies the following requirements: (i) $\tilde{B}_{\mu\nu}$ is a linear function that is first-order in $B_{\mu\nu}$ and second-order in $N_{\mu}$; (ii) $\tilde{B}_{\mu\nu}$ has mass dimension one in $4D$ spacetime; (iii) $\tilde{B}_{\mu\nu}$ transforms by $\tilde{B}_{\mu\nu}\rightarrow\tilde{B}_{\mu\nu}+\partial_{\mu}\tilde{\Sigma}_{\nu}-\partial_{\nu}\tilde{\Sigma}_{\mu}$ when $B_{\mu\nu}$ changes by $B_{\mu\nu}\rightarrow B_{\mu\nu}+\tilde{\partial}_{\mu}\Sigma_{\nu}-\tilde{\partial}_{\nu}\Sigma_{\mu}$.   

After imposing these requirements, we arrive at the $\tilde{B}$-ansatz given by
\begin{equation}
\tilde{B}_{\mu\nu}=B_{\mu\nu}-\frac{m^{2}}{2}\left(N_{\mu}N^{\alpha}B_{\alpha\nu}-N_{\nu}N^{\alpha}B_{\alpha\mu}\right),
\end{equation}
and it changes under the gauge transformation $B_{\mu\nu}\rightarrow B_{\mu\nu}+\tilde{\partial}_{\mu}\Sigma_{\nu}-\tilde{\partial}_{\nu}\Sigma_{\mu}$ as follows:
\begin{equation}
\tilde{B}_{\mu\nu}\rightarrow\tilde{B}_{\mu\nu}+\partial_{\mu}\left(\Sigma_{\nu}-\frac{m^{2}}{2}N_{\nu}(N\cdot\Sigma)\right)-\partial_{\nu}\left(\Sigma_{\mu}-\frac{m^{2}}{2}N_{\mu}(N\cdot\Sigma)\right).
\end{equation}

It is worth noting that the gauge parameter $\tilde{\Sigma}_{\mu}=\Sigma_{\mu}-\frac{m^{2}}{2}N_{\mu}(N\cdot\Sigma)$ has the same form as $\tilde{A}_{\mu}$ in Eq. (\ref{Atilde}), which was obtained in the Maxwell case. This is expected since $\Sigma_{\mu}$ has the additional gauge symmetry (\ref{gaugeSigma}), similar to the $A_{\mu}$ field. Additionally, it is interesting to note that for a 0-form field, our prescription implies that $\tilde{\phi}=\phi-\frac{m^{2}}{2}N^{\mu}N_{\mu}\phi=\phi$, because $N^{2}=0$.

Once we have found $\tilde{B}_{\mu\nu}$, we can define the tensor
\begin{equation}
\mathcal{H_{\mu\nu\alpha}=}\partial_{\mu}\tilde{B}_{\nu\alpha}+\partial_{\alpha}\tilde{B}_{\mu\nu}+\partial_{\nu}\tilde{B}_{\alpha\mu},
\end{equation}whose explicit form is given by:
\begin{equation}
\mathcal{H}_{\mu\nu\alpha}\mathcal{=}H_{\mu\nu\alpha}+\frac{1}{2}m^{2}\left[N_{\mu}N^{\sigma}\left(\partial_{\nu}B_{\sigma\alpha}-\partial_{\alpha}B_{\sigma\nu}\right)+N_{\nu}N^{\sigma}\left(\partial_{\alpha}B_{\sigma\mu}-\partial_{\mu}B_{\sigma\alpha}\right)+N_{\alpha}N^{\sigma}\left(\partial_{\mu}B_{\sigma\nu}-\partial_{\nu}B_{\sigma\mu}\right)\right].
\end{equation}

%It is noteworthy to comment that a similar result was obtained in Ref. \cite{Upadhyay:2015yyt}, but by a different approach. 

Also, the $SIM(2)$-covariant field strength tensor $\tilde{H}_{\mu\nu\alpha}$ can be defined as
\begin{equation}
\tilde{H}_{\mu\nu\alpha}\equiv\tilde{\partial}_{\mu}B_{\nu\alpha}+\tilde{\partial}_{\alpha}B_{\mu\nu}+\tilde{\partial}_{\nu}B_{\alpha\mu}.
\end{equation}
Taking the difference between the two kinds of field strengths $\tilde{H}_{\mu\nu\alpha}-\mathcal{H}_{\mu\nu\alpha}$, we obtain
\begin{equation}
\tilde{H}_{\mu\nu\alpha}-\mathcal{H}_{\mu\nu\alpha}=\frac{1}{2}m^{2}\left[N_{\mu}N^{\sigma}H_{\sigma\nu\alpha}+N_{\nu}N^{\sigma}H_{\sigma\alpha\mu}+N_{\alpha}N^{\sigma}H_{\sigma\mu\nu}\right].
\end{equation}

Furthermore, we can rewrite $\tilde{H}_{\mu\nu\alpha}$ solely in terms of $\mathcal{H}_{\mu\nu\alpha}$, thereby guaranteeing the invariance of $\tilde{H}_{\mu\nu\alpha}$ under both $SIM(2)$ and the wiggle gauge transformations, as required in condition (iii). To this end, it is easy to check the following identity:
\begin{equation}
N_{\mu}N^{\sigma}H_{\sigma\nu\alpha}+N_{\nu}N^{\sigma}H_{\sigma\alpha\mu}+N_{\alpha}N^{\sigma}H_{\sigma\mu\nu}=N_{\mu}N^{\sigma}\mathcal{H}_{\sigma\nu\alpha}+N_{\nu}N^{\sigma}\mathcal{H}_{\sigma\alpha\mu}+N_{\alpha}N^{\sigma}\mathcal{H}_{\sigma\mu\nu},
\end{equation}
where using the properties of the operator $N_{\mu}$ we have that \cite{Cheon:2009zx} 
\begin{equation}
N\cdot N=0,\ \ N\cdot\partial=1,\ \ \left[N^{\mu},N^{\nu}\right]=\left[N^{\mu},\partial^{\nu}\right]=0,
\end{equation} 
and the integration by parts rule holds:
\begin{equation}
\int d^{4}xf(x)\left(N^{\mu}g(x)\right)=-\int d^{4}x\left(N^{\mu}f(x)\right)g(x).
\end{equation}
Besides, it would also be consistent to set $N^{\mu}\phi(x)\equiv0$ if $\phi$ is a constant. With all the above results, the $SIM(2)$-covariant tensor $\tilde{H}_{\mu\nu\alpha}$ can be cast as
\begin{equation}
\tilde{H}_{\mu\nu\alpha}=\mathcal{H}_{\mu\nu\alpha}+\frac{1}{2}m^{2}\left[N_{\mu}N^{\sigma}\mathcal{H}_{\sigma\nu\alpha}+N_{\nu}N^{\sigma}\mathcal{H}_{\sigma\alpha\mu}+N_{\alpha}N^{\sigma}\mathcal{H}_{\sigma\mu\nu}\right] \ , \label{Htilde1}
\end{equation}
which is a natural generalization of the relation (\ref{Ftilda2}). 

Finally, the $SIM(2)$-invariant action of the Kalb-Ramond field $B_{\mu\nu}$ is represented by 
\begin{equation}
S=\int d^{4}x\left[-\frac{1}{12}\tilde{H}_{\mu\nu\alpha}\tilde{H}^{\mu\nu\alpha}\right],\label{Htilde2}
\end{equation}
and with help the of Eq. (\ref{Htilde1}) we can write it as 
\begin{equation}
S=\int d^{4}x\left[-\frac{1}{12}H_{\mu\nu\alpha}H^{\mu\nu\alpha}+\frac{1}{4}m^{2}n^{\alpha}\left(\frac{1}{n\cdot\partial}H_{\alpha\mu\nu}\right)n_{\sigma}\left(\frac{1}{n\cdot\partial}H^{\sigma\mu\nu}\right)\right],\label{Htilde3}
\end{equation}
where we performed an additional change of field variables $B_{\mu\nu}\rightarrow B_{\mu\nu}+\frac{m^{2}}{2}\left(N_{\mu}N^{\alpha}B_{\alpha\nu}-N_{\nu}N^{\alpha}B_{\alpha\mu}\right)$. Therefore, similar to the Maxwell case, the $SIM(2)$-modified Kalb-Ramond action (\ref{Htilde2}) is invariant under the standard gauge transformation (\ref{gaugeKR}).

The equation of motion follows from the action (\ref{Htilde3}) by varying with respect to $B_{\nu\alpha}$, Explicitly, we find
\begin{equation}
\partial_{\mu}H^{\mu\nu\alpha}+m^{2}\left(N_{\sigma}H^{\sigma\nu\alpha}+N^{\nu}N_{\sigma}\partial_{\mu}H^{\sigma\alpha\mu}+N^{\alpha}N_{\sigma}\partial_{\mu}H^{\sigma\mu\nu}\right)=0.\label{EoMB}
\end{equation}
By contracting Eq. (\ref{EoMB}) with $N_{\nu}$, we obtain the following constraint:
\begin{equation}
N_{\nu}\partial_{\mu}H^{\mu\nu\alpha}=0.\label{ct}
\end{equation}
Inserting this constraint back into the equation of motion, we find
\begin{equation}
\partial_{\mu}H^{\mu\nu\alpha}+m^{2}N_{\sigma}H^{\sigma\nu\alpha}=0.\label{EoMB2}
\end{equation}
To find the physical modes, we must fix the gauge freedom. We can chose, analogous
to the Maxwell case, the Lorentz gauge
\begin{equation}
\partial_{\mu}B^{\mu\nu}=0.\label{GaugeB}
\end{equation}Then the equation of motion (\ref{EoMB2}) and the constraint (\ref{ct}) become, respectively:
\begin{equation}
\left(\Box+m^{2}\right)B^{\nu\alpha}+m^{2}\left(N_{\sigma}\partial^{\alpha}B^{\sigma\nu}+N_{\sigma}\partial^{\nu}B^{\alpha\sigma}\right)=0,\label{EoMB3}
\end{equation}and
\begin{equation}
\Box N_{\nu}B^{\nu\alpha}=0.\label{CondNB1}
\end{equation} 

The form of Eq. (\ref{EoMB3}) still contains redundant degrees of freedom. Indeed,  the gauge condition (\ref{GaugeB}) is insufficient to fix the gauge freedom completely since we can construct a solution $B'^{\nu\alpha}=B^{\nu\alpha}+\partial^{\nu}\Sigma^{\alpha}-\partial^{\alpha}\Sigma^{\nu}$, which preserves the Lorentz gauge (\ref{GaugeB}) and satisfies the equation of motion (\ref{EoMB3}). So, we can impose an additional condition on the field $B'^{\nu\alpha}$, namely, 
\begin{equation}
N_{\nu}B'^{\nu\alpha}=0,\label{CondNB2}
\end{equation}
by choosing the gauge parameter $\Sigma^{\alpha}$ as
\begin{equation}
\Sigma^{\alpha}-\partial^{\alpha}N_{\nu}\Sigma^{\nu}=-N_{\nu}B^{\nu\alpha}.
\end{equation}However, the last relation is invariant under the residual gauge symmetry (\ref{gaugeSigma}), and we can use this fact to impose the condition 
\begin{equation}
N_{\nu}\Sigma^{\nu}=0,
\end{equation}by fixing the scalar gauge parameter as $\phi=N_{\nu}\Sigma'^{\nu}$. Thus, the gauge parameter is given by
\begin{equation}
\Sigma^{\alpha}=-N_{\nu}B^{\nu\alpha}.\label{SigmaNB}
\end{equation}
Now, we can show that the Lorentz gauge condition is valid to $B'^{\nu\alpha}$. From (\ref{GaugeB}) we find 
\begin{equation}
\partial_{\nu}B'^{\nu\alpha}=\Box\Sigma^{\alpha}-\partial^{\alpha}\partial_{\nu}\Sigma^{\nu}.\label{Bprimegauge}
\end{equation}On the other hand, the constraint condition (\ref{CondNB1}) implies that 
\begin{equation}
\Box N_{\nu}B'^{\nu\alpha}=\Box\Sigma^{\alpha}=0,
\end{equation}where we used the relations $N_{\nu}B'^{\nu\alpha}=0$ and $N_{\nu}\Sigma^{\nu}=0$. Immediately, from (\ref{Bprimegauge}) it follows that
\begin{equation}
\partial_{\nu}B'^{\nu\alpha}=-\partial^{\alpha}\partial_{\nu}\Sigma^{\nu},
\end{equation}
and by Eq. (\ref{SigmaNB}), we obtain the claimed result
\begin{equation}
\partial_{\nu}B'^{\nu\alpha}=\partial^{\alpha}\partial_{\nu}N_{\mu}B^{\mu\nu}=0,
\end{equation}where in the last step we use again the Lorentz gauge $\partial_{\mu}B^{\mu\nu}=0$. 

Finally, by applying the subsidiary condition $N_{\nu}B^{\nu\alpha}=0$ to the equation of motion (\ref{EoMB3}), it takes the simple form
\begin{equation}
\left(\Box+m^{2}\right)B^{\nu\alpha}=0,\label{KG}
\end{equation}which represents a wave equation for a particle of mass $m$. At the end, the Kalb-Ramond field in VSR satisfies the standard
Klein-Gordon equation under two gauge conditions 
\begin{equation}
\partial_{\nu}B^{\nu\alpha}=0,\ \ \ N_{\nu}B^{\nu\alpha}=0.\label{GC}
\end{equation}

The general solution to Eq. (\ref{KG}) takes the form:
\begin{equation}
B^{\mu\nu}(x)=\int\frac{d^{3}{\bf p}}{\sqrt{(2\pi)^{3}2\omega_{{\bf p}}}}\left[B^{\mu\nu}({\bf p})e^{ip\cdot x}+B^{\mu\nu*}({\bf p})e^{-ip\cdot x}\right],\label{SolKBR}
\end{equation}where $\omega_{{\bf p}}=\sqrt{{\bf p^{2}}+m^{2}}$, and the associated 4-momenta $p^{\mu}$ are on shell such that $p^{\mu}=(\omega_{{\bf p}},{\bf p})$. The Fourier coefficients $B^{\mu\nu}({\bf p})$ can be expanded over a basis of polarization antisymmetric 2-tensors, labeled by $\lambda=1,\cdots,6$:
\begin{equation}
B^{\mu\nu}({\bf p})=a_{{\bf p},\lambda}\epsilon^{\mu\nu}({\bf p},\lambda).
\end{equation}

To find the physical polarization states, it is convenient to analyze the solution (\ref{SolKBR}) in the rest frame where $k^{\mu}=(m,{\bf 0})$. The solution for a general $p^{\mu}$ can then be obtained by applying a VSR boost, i.e., $p^{\mu}=L(p)^{\mu}_{\ \nu} k^{\nu}$, with
\begin{align}
L(p) & =T_{1}(\beta_{1})T_{2}(\beta_{2})L_{3}(\xi)\nonumber \\
 & =e^{i\beta_{1}\mathcal{T}^{1}}e^{i\beta_{2}\mathcal{T}^{2}}e^{i\xi\mathcal{K}^{3}},
\end{align}where $\mathcal{T}^{1}$, $\mathcal{T}^{2}$, and $\mathcal{K}^{3}$ are the generators of the $SIM(2)$ group in the vector representation \cite{Lee:2015tcc}.

In the rest frame, the gauge conditions (\ref{GC}) become
\begin{equation}
k_{\mu}\epsilon^{\mu\nu}({\bf k},\lambda)=0,\ \ \ \ n_{\mu}\epsilon^{\mu\nu}({\bf k},\lambda)=0.
\end{equation}The first condition implies that $\epsilon^{0i}({\bf k},\lambda)=0$ with $i=1,2,3$, which eliminates three polarizations. The second condition gives $\epsilon^{3j}({\bf k},\lambda)=0$ with $j=1,2$, which eliminates two more polarizations. Therefore, there is only one non-zero polarization $\epsilon^{12}({\bf k},\lambda)$, which means that the free Kalb-Ramond field in VSR has only one degree of freedom, equivalent to a single massive scalar field.

%%%%%%%%%%%%%%%%%%%%%%%%%%%%%%%%%%%%%%%%%%%%%%%%%%%%%%%%%%%%%%%%%%%%%%%%%%%%%%%%
\section{Maxwell-Kalb-Ramond vacuum polarization in VSR \label{sec4}}

In this section, we calculate the effective action for the case of a fermion field interacting with Maxwell and Kalb-Ramond fields within the context of VSR, which has been developed in the preceding sections. As we proceed, we will obtain exact solutions to the one-loop vacuum polarization amplitudes involving the external gauge fields.

Let us start by recalling that Kalb-Ramond quantum electrodynamics is a $U(1)$ gauge theory that involves a 2-form gauge potential $B_{\mu\nu}(x)$ and a string matter field $\psi(x(\sigma))$ serving as the source for $B_{\mu\nu}(x)$ \cite{Kalb:1974yc}. Considering the complexities inherent in string theory-based systems, let us focus on exploring a simplified scenario within the framework of VSR. Specifically, we investigate an interaction model involving a point-like fermion field and the Maxwell and Kalb-Ramond fields in four-dimensional Minkowski spacetime. As we will see below, this type of interaction is only viable if $\psi$ does not carry any Kalb-Ramond charge and couples nonminimally with it \cite{HariDass:2001dp}.

We consider $SIM(2)$-covariant gauge theories under the $U(1)$ gauge transformations
\begin{align}
B_{\mu\nu} & \rightarrow B_{\mu\nu}+\overset{\nsim}{\partial}_{\mu}\Sigma_{\nu}-\overset{\nsim}{\partial}_{\nu}\Sigma_{\mu},\\
A_{\mu} & \to A_{\mu}+\tilde{\partial}_{\mu}\Lambda,\\
\psi & \rightarrow\exp\left\{ ie\Lambda\right\} \psi,
\end{align}where we have defined the wiggle operators as
\begin{equation}
\tilde{\partial}_{\mu}=\partial_{\mu}+\frac{1}{2}\frac{m_{A}^{2}}{n\cdot\partial}n_{\mu},\ \ \ \ \text{and}\ \ \ \ \overset{\nsim}{\partial}_{\mu}=\partial_{\mu}+\frac{1}{2}\frac{m_{B}^{2}}{n\cdot\partial}n_{\mu},
\end{equation}where $m_{A}$ and $m_{B}$ represent the VSR-mass associated with the Maxwell and Kalb-Ramond fields, respectively.

The simplest Lagrangian density that can be constructed, invariant under the aforementioned gauge transformations, is given by \cite{HariDass:2001dp}:
\begin{align}
\mathcal{L} & =-\frac{1}{4}\tilde{F}_{\mu\nu}\tilde{F}^{\mu\nu}-\frac{1}{12}\overset{\nsim}{H}_{\mu\nu\alpha}\overset{\nsim}{H}^{\mu\nu\alpha}\nonumber\\
 & +\bar{\psi}\left(i\cancel{D}+i\frac{m_{\psi}^{2}}{2}\frac{\cancel{n}}{n\cdot D}+\frac{1}{12}\frac{g}{m}\sigma_{\mu\nu\lambda}\overset{\nsim}{H}^{\mu\nu\lambda}-m\right)\psi,\label{LMKR}
\end{align}where $m$ represents the usual fermion mass, $m_{\psi}$ the VSR-mass associated to the $\psi$ field, and $g$ is a coupling constant with mass dimension $[g]=M^{-2}$ (in natural units). Also, the operator $D_{\mu}$ denotes the standard covariant derivative, given by
\begin{equation}
D_{\mu}=\partial_{\mu}-ieA_{\mu},
\end{equation}and $\sigma_{\mu\nu\lambda}$ represents the fully antisymmetrized product of two gamma matrices normalized to unit strength, defined as
\begin{equation}
\sigma_{\mu\nu\lambda}=i\epsilon_{\mu\nu\lambda\alpha}\gamma_{5}\gamma^{\alpha},
\end{equation}
where $\epsilon_{\mu\nu\lambda\alpha}$ is the Levi-Civita symbol. It satisfies the commutation relation
\begin{equation}
[\sigma_{\mu\nu\lambda},\gamma_{\sigma}]=2i\epsilon_{\mu\nu\lambda\sigma}\gamma_{5}.
\end{equation}

It is important to note that the Lagrangian density (\ref{LMKR}) involves a nonminimal coupling of the Kalb-Ramond field to point-like fermions. We could interpret this kind of coupling as similar to that involving neutral particles, such as the neutron, interacting with the electromagnetic field through their anomalous magnetic moments.

In order to determine the effective Lagrangian resulting from fermionic vacuum polarization at the one-loop level, we consider the generating functional defined as
\begin{equation}
e^{iS_{\text{eff}}[A,B]}=\mathcal{N}\int\mathcal{D}\bar{\psi}\mathcal{D}\psi e^{i\int d^{4}x\left[-\frac{1}{4}\tilde{F}_{\mu\nu}\tilde{F}^{\mu\nu}-\frac{1}{12}\overset{\nsim}{H}_{\mu\nu\alpha}\overset{\nsim}{H}^{\mu\nu\alpha}+\bar{\psi}\left(i\cancel{D}+i\frac{m_{\psi}^{2}}{2}\frac{\cancel{n}}{n\cdot D}+\frac{1}{12}\frac{g}{m}\sigma_{\mu\nu\lambda}\overset{\nsim}{H}^{\mu\nu\lambda}-m\right)\psi\right]},
\end{equation}where $\mathcal{N}$ is a normalization constant which will be used to
absorb field-independent factors.

By performing the fermionic integration, we obtain (up to a field-independent factor that can be absorbed in the normalization):
\begin{align}
S_{\text{eff}}[A,B] & =\int d^{4}x\left[-\frac{1}{4}\tilde{F}_{\mu\nu}\tilde{F}^{\mu\nu}-\frac{1}{12}\overset{\nsim}{H}_{\mu\nu\alpha}\overset{\nsim}{H}^{\mu\nu\alpha}\right]\nonumber\\
 & -i\mbox{Tr}\ensuremath{\ln\left[i\cancel{D}+i\frac{m_{\psi}^{2}}{2}\frac{\cancel{n}}{n\cdot D}+\frac{1}{12}\frac{g}{m}\sigma_{\mu\nu\lambda}\overset{\nsim}{H}^{\mu\nu\lambda}-m\right]},\label{Seef1}
\end{align}where $\text{Tr}$ stands for the trace over Dirac matrices as well as the trace over the integration in coordinate space. At this point, it is worth mentioning that the Lorentz covariant calculation of vacuum polarization for the model under study was carried out in Ref. \cite{HariDass:2001dp} for the simplified case of constant fields using Schwinger's approach. In this work, we extend this calculation to the $SIM(2)$-covariant case using Feynman techniques without the restriction to constant external gauge fields. Furthermore, we can verify whether our results can reproduce those obtained in the literature by taking the limit $m_{A,B,\psi}\rightarrow 0$ and assuming constant $F$ and $H$ fields.

To evaluate the trace in Eq. (\ref{Seef1}), we notice that the term $1/(n \cdot D)$ is a non-local operator that depends on the vector gauge field $A_{\mu}$. Since our focus is on perturbative computations, we treat the non-local term as a perturbative expansion in the fermionic Lagrangian density. This is accomplished by using the following matrix identity:
\begin{equation}
\frac{1}{A+B}=\frac{1}{A}-\frac{1}{A}B\frac{1}{A+B}=\frac{1}{A}-\frac{1}{A}B\frac{1}{A}+\frac{1}{A}B\frac{1}{A}B\frac{1}{A+B}=\cdots
\end{equation}

Therefore, we can obtain the corresponding new types of vertices involving more than one external gauge vector field from the perturbative Lagrangian density:
\begin{align}
\mathcal{L}_{\text{fermion}} & =\bar{\psi}\left[\left(i\cancel{\partial}+i\frac{m_{\psi}^{2}}{2}\frac{\cancel{n}}{n\cdot\partial}-m\right)\right.\nonumber\\
 & +e\left(\cancel{A}-\frac{m_{\psi}^{2}}{2}\frac{\cancel{n}}{n\cdot\partial}(n\cdot A)\frac{1}{n\cdot\partial}\right)+\frac{1}{12}\frac{g}{m}\sigma_{\mu\nu\lambda}\overset{\nsim}{H}^{\mu\nu\lambda}\nonumber\\
 & -ie^{2}\left(\frac{m_{\psi}^{2}}{2}\frac{\cancel{n}}{n\cdot\partial}(n\cdot A)\frac{1}{n\cdot\partial}(n\cdot A)\frac{1}{n\cdot\partial}\right)\nonumber\\
 & \left.e^{3}\left(\frac{m_{\psi}^{2}}{2}\frac{\cancel{n}}{n\cdot\partial}(n\cdot A)\frac{1}{n\cdot\partial}(n\cdot A)\frac{1}{n\cdot\partial}(n\cdot A)\frac{1}{n\cdot\partial}\right)+\cdots\right]\psi.\label{LpsiVSR}
\end{align}

Thus, we can write the nontrivial part of the effective action as 
\begin{equation}
S_{\text{eff}}^{(n)}[A,B]=i\mbox{Tr}\sum_{n=1}^{\infty}\frac{1}{n}\left[\frac{i}{i\cancel{\partial}+i\frac{m_{\psi}^{2}}{2}\frac{\cancel{n}}{n\cdot\partial}-m}i\hat{\mathcal{O}}\right]^{n},\label{Seffn}
\end{equation}where the operator $\hat{\mathcal{O}}$ can be determined from Eq. (\ref{LpsiVSR}).

\begin{figure}[h!]
\begin{center}
\begin{tabular}{cc}
\includegraphics[scale=0.15]{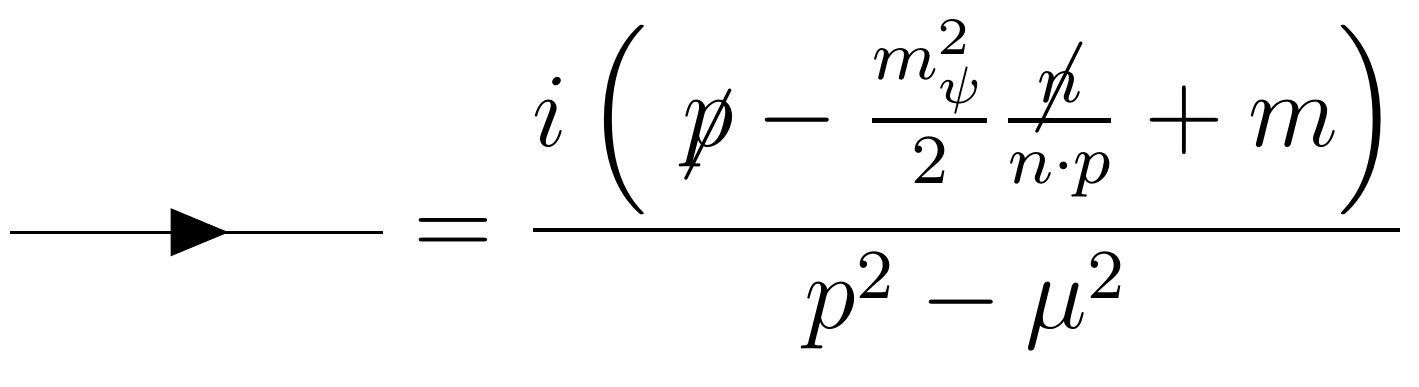}
\hspace{1cm}
\includegraphics[scale=0.15]{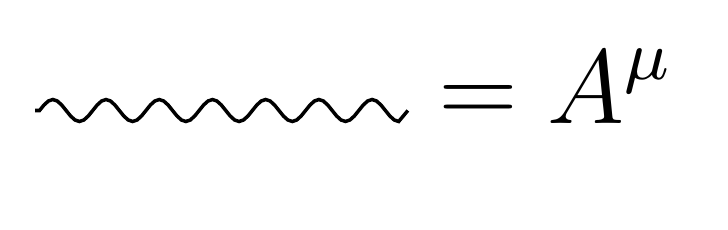}
\hspace{1cm} 
\includegraphics[scale=0.15]{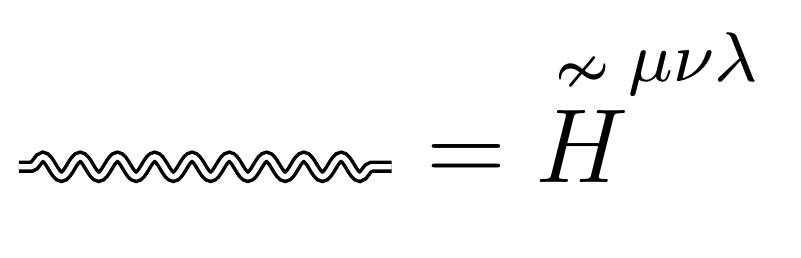}
\end{tabular}
\end{center}
\caption{Feynman rules. Continuous, wave, and double wave lines represent
the fermion propagator, the gauge vector $A^{\mu}$, and the field strength $\overset{\nsim}{H}^{\mu\nu\lambda}$, respectively.\label{Fig1}}
\end{figure}

\begin{figure}[h!]
\begin{center}
\begin{tabular}{cc}
\includegraphics[scale=0.15]{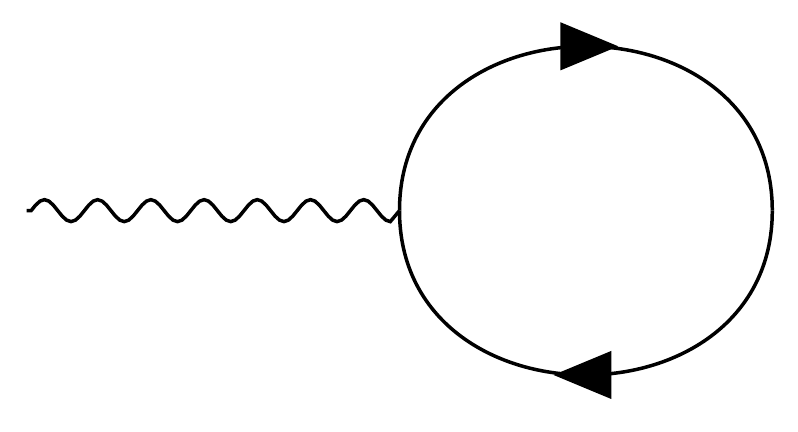}
\hspace{1cm}
\includegraphics[scale=0.15]{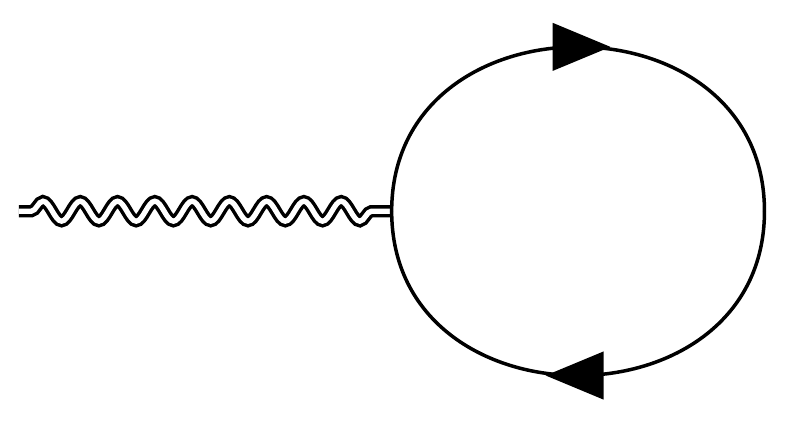}
\hspace{1cm}
\includegraphics[scale=0.15]{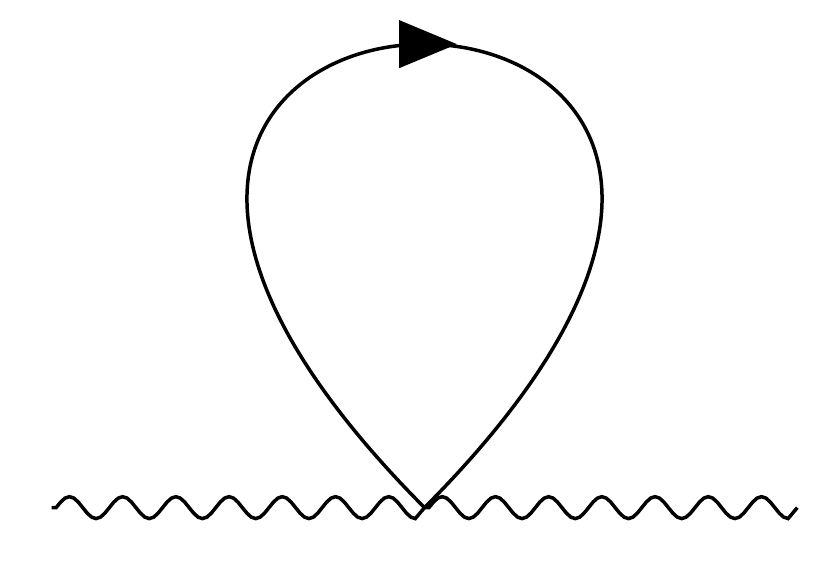}\\
\hspace{0.7cm} (a) \hspace{3.6cm} (b) \hspace{3.1cm} (c)  
\end{tabular}
\end{center}
\caption{The contributions corresponding to $n=1$ include: (a) Tadpole-$A$, (b) Tadpole-$H$, and (c) Vacuum polarization $AA$.\label{Fig2}}
\end{figure}

\begin{figure}[h!]
\begin{center}
\begin{tabular}{cc}
\includegraphics[scale=0.15]{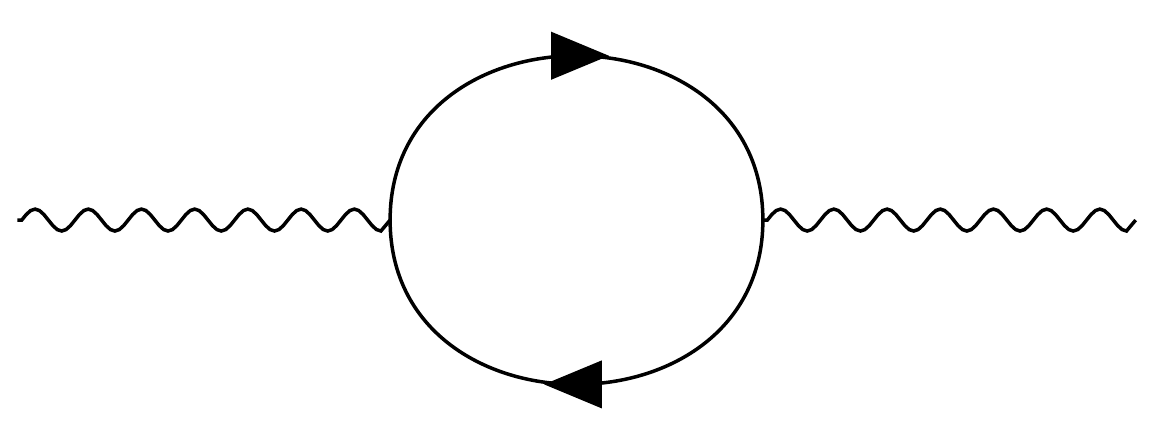}
\hspace{1cm}
\includegraphics[scale=0.15]{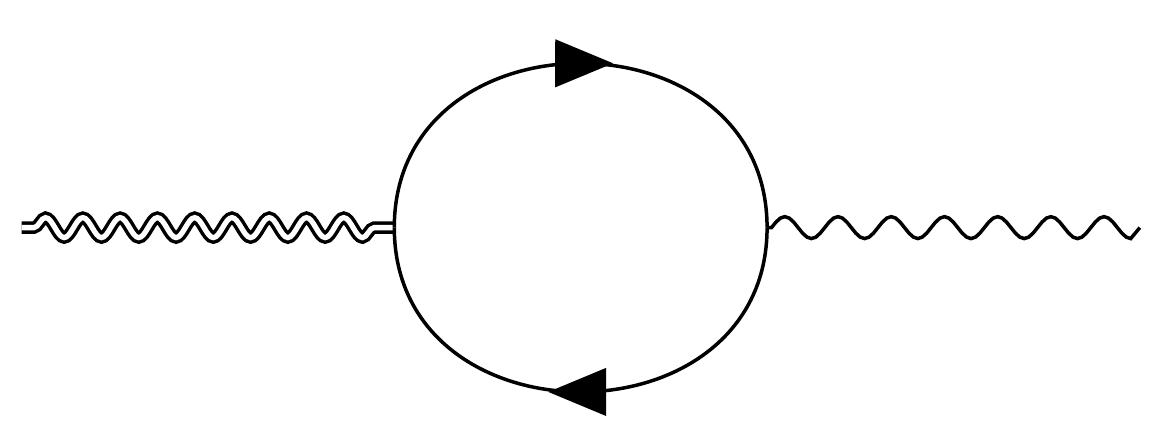}\\
(a)\hspace{5.3cm} (b)\\
\includegraphics[scale=0.15]{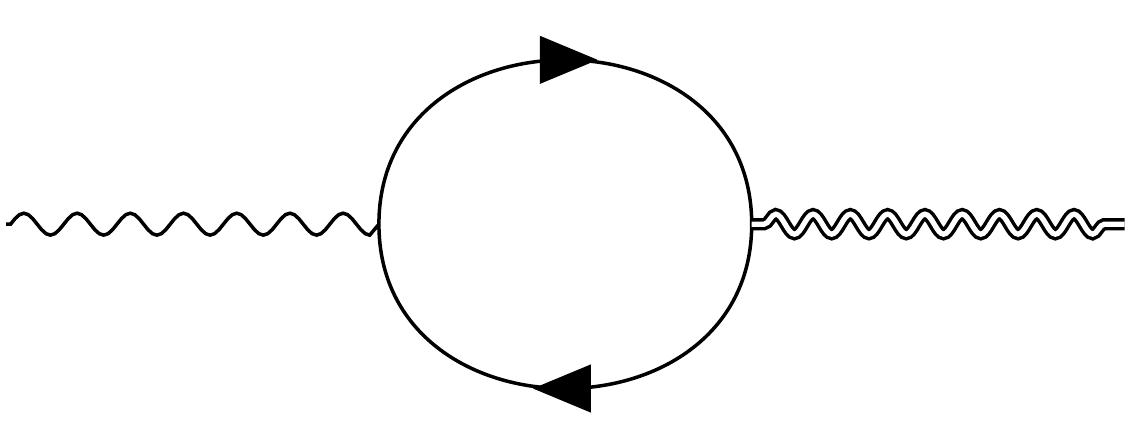}
\hspace{1cm}
\includegraphics[scale=0.15]{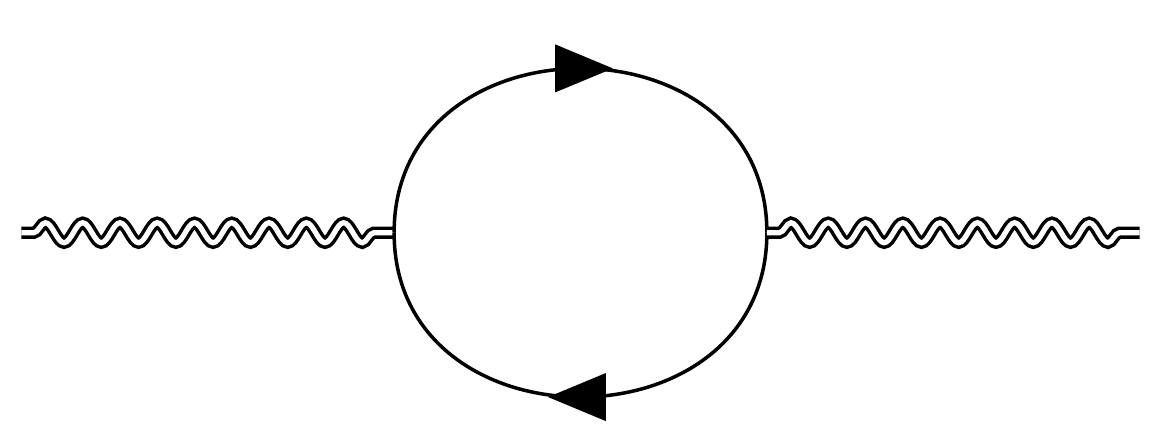}\\
(c) \hspace{5.3cm} (d)   
\end{tabular}
\end{center}
\caption{The contributions corresponding to $n=2$ include: (a) Vacuum polarization $AA$, (b) Vacuum polarization $HA$, (c) Vacuum polarization $AH$ and (d) Vacuum polarization $HH$.\label{Fig3}}
\end{figure}

The formal contributions of this formula will give rise to the $n$-point vertex functions of the fields $A_{\mu}$ and $B_{\mu\nu}$. At this point, a graphical representation may be helpful. Following the conventions depicted in Fig. \ref{Fig1}, the contributions to the tadpole and the self-energy are illustrated in Figs. \ref{Fig2} and \ref{Fig3} up to one-loop order.

%\revision{The contributions to the vacuum polarization come from the $n=1$ and $n=2$, i.e., second-order contributions in $e$ and $g$. Hence, after the trace evaluation over the space coordinate we obtain}

For $n=1$ the expression (\ref{Seffn}) gives rise to three contributions (\ref{Seffn1}) up to second order in the coupling constant $e$, as graphically indicated in Fig. \ref{Fig2}. 
\begin{equation}
S_{\text{eff}}^{(1)}=S_{\text{eff}}^{(1,a)}+S_{\text{eff}}^{(1,b)}+S_{\text{eff}}^{(1,c)}.\label{Seffn1}
\end{equation}

We found that the tadpole diagrams 1(a) and 1(b) vanish, as expected by Furry's theorem \cite{Bufalo:2020cst}. The remaining contribution, i.e., the self-energy 1(c),
gives
\begin{equation}
S_{\text{eff}}^{(1,c)}=i\mbox{Tr}\left[\frac{i}{i\cancel{\partial}+i\frac{m_{\psi}^{2}}{2}\frac{\cancel{n}}{n\cdot\partial}-m}e^{2}\left(\frac{m_{\psi}^{2}}{2}\frac{\cancel{n}}{n\cdot\partial}(n\cdot A)\frac{1}{n\cdot\partial}(n\cdot A)\frac{1}{n\cdot\partial}\right)\right],
\end{equation}and we can write it in momentum space. The result is
\begin{equation}
S_{\text{eff}}^{(1,c)}=\frac{i}{2}\int\frac{d^{4}q}{(2\pi)^{4}}\Pi_{\mu\nu}^{(1,c)}A^{\mu}(q)A^{\nu}(-q),
\end{equation}
where 
\begin{equation}
\Pi_{\mu\nu}^{(1,c)}=-ie^{2}m_{\psi}^{2}\mbox{Tr}\int\frac{d^{4}p}{(2\pi)^{4}}\frac{i(\cancel{\tilde{p}}+m)}{p^{2}-\mu^{2}}\frac{\cancel{n}n_{\mu}n_{\nu}}{(n\cdot p)^{2}n\cdot u},\label{Pi1c}
\end{equation}with $u=p+q$, and $\mu^{2}=m_{\psi}^{2}+m^{2}$ representing the modified fermion mass. Besides, we define the wiggle momentum by
\begin{equation}
\tilde{p}_{\mu}=p_{\mu}-\frac{m_{\psi}^{2}}{2}\frac{n_{\mu}}{n\cdot p}.
\end{equation}

After calculating the traces of the Dirac matrices in four dimensions, the expression (\ref{Pi1c}) yields
\begin{equation}
\Pi_{\mu\nu}^{(1,c)}=\int\frac{d^{4}p}{(2\pi)^{4}}\frac{4e^{2}m_{\psi}^{2}n_{\mu}n_{\nu}}{(p^{2}-\mu^{2})(n.p)(n.u)}.\label{Pi1c2}
\end{equation}
This integral is ultraviolet divergent and also exhibits an infrared divergence when $(n \cdot p)=0$. 

To deal with these types of divergences in VSR, we use the Mandelstan-Leibbrant prescription \cite{Leibbrandt:1983pj,Leibbrandt:1987qv,LeibbrandtBook}
\begin{equation}
\frac{1}{n\cdot p}=\lim_{\epsilon\rightarrow0}\frac{(\bar{n}\cdot p)}{(n\cdot p)(\bar{n}\cdot p)+i\epsilon},\label{MLpre}
\end{equation} where $\bar{n}_{\mu}$ is a extra null vector which obeys $(n\cdot\bar{n})=1$.  
Moreover, we employ a useful decomposition formula 
\begin{equation}
\frac{1}{n\cdot(p+q_{i})n\cdot(p+q_{j})}=\frac{1}{n\cdot(q_{i}-q_{j})}\left(\frac{1}{n\cdot(p+q_{j})}-\frac{1}{n\cdot(p+q_{i})}\right),
\end{equation}to isolate factors as $1/(n\cdot p)$ in each momentum integration. The resulting loop integrals can be evaluated using the formulas introduced by Alfaro in Ref. \cite{Alfaro:2016pjw}. Here, we will quote the most basic of them:
\begin{align}
\int dp\frac{1}{\left[p^{2}+2p\cdot q-m^{2}\right]^{a}}\frac{1}{\left(n\cdot p\right)^{b}} & =(-1)^{a+b}i\pi^{\omega}(-2)^{b}\frac{\Gamma(a+b-\omega)}{\Gamma(a)\Gamma(b)}\left(\bar{n}\cdot q\right)^{b}\nonumber \\
 & \times\int_{0}^{1}dtt^{b-1}\frac{1}{\left(m^{2}+q^{2}-2(n\cdot q)(\bar{n}\cdot q)t\right)^{a+b-\omega}},\label{IntegrateFormule}
\end{align}
where $dp$ is the integration measure in $d$-dimensional space with $\omega=d/2$.

Following the calculation procedure mentioned above to evaluate the integral in Eq. (\ref{Pi1c2}), we obtain:
\begin{equation}
\Pi_{\mu\nu}^{(1,c)}=-i\frac{e^{2}m_{\psi}^{2}}{n\cdot q}\int_{0}^{1}dt\left[\frac{\Gamma(2-\omega)(\bar{n}\cdot q)n_{\mu}n_{\nu}}{2^{2\omega-3}\pi^{\omega}\left(\mu^{2}-2t(n\cdot q)(\bar{n}\cdot q)\right)^{2-\omega}}\right].\label{Pi1c3}
\end{equation}

As we can note, the Mandelstam-Leibbrandt prescription (\ref{MLpre}) adopted in the formula for the loop integral (\ref{IntegrateFormule}) introduces a new null vector $\bar{n}_{\mu}$, which {\it a priori} could break the $SIM(2)$ symmetry. To preserve the $SIM(2)$ covariance in this calculation, Alfaro proposes in Ref. \cite{Alfaro:2017umk}  to fix the vector $\bar{n}_{\mu}$ as a linear combination of the original null vector $n_{\mu}$ and the external momentum of the diagram. By imposing certain conditions, such as reality, appropriate scaling $(n, \bar{n})\rightarrow(\lambda n, \lambda^{-1}\bar{n})$, and being dimensionless \cite{Bufalo:2020cst, Haghgouyan:2022dvq}, the vector $\bar{n}_{\mu}$ can be expressed in the following form:
\begin{equation}
\bar{n}_{\mu}=\frac{q_{\mu}}{n\cdot q}-\frac{q^{2}n_{\mu}}{2(n\cdot q)^{2}},
\end{equation}which leads to $\bar{n}\cdot q=\frac{q^{2}}{2(n\cdot q)}$. Hence, we can substitute this result into Eq. (\ref{Pi1c3}), and after integrating over the variable $t$, we find
\begin{equation}
\Pi_{\mu\nu}^{(1,c)}=-\frac{ie^{2}m_{\psi}^{2}}{4\pi^{2}}\left\{ \frac{q^{2}n_{\mu}n_{\nu}}{(2-\omega)(n\cdot q)^{2}}+\frac{n_{\mu}n_{\nu}}{(n\cdot q)^{2}}\left[\mu^{2}\ln\left(1-\frac{q^{2}}{\mu^{2}}\right)+q^{2}\left(1-\gamma_{\text{E}}-\ln\left(\frac{\mu^{2}-q^{2}}{4\pi}\right)\right)\right]\right\}, 
\end{equation}where we have performed an expansion around $\omega=2$ and retained only the $1/(2-\omega)$ pole and the finite terms when $\omega\rightarrow2^{+}$. We note that this result is exclusively an effect of the VSR and goes to zero when we take the limit $m_{\psi}^{2}\rightarrow 0$.

For $n=2$ the expression (\ref{Seffn}) yields four 2-point amplitudes up to one-loop order 
\begin{equation}
S_{\text{eff}}^{(2)}=S_{\text{eff}}^{(2,a)}+S_{\text{eff}}^{(2,b)}+S_{\text{eff}}^{(2,c)}+S_{\text{eff}}^{(2,d)},
\end{equation}as depicted in Fig. \ref{Fig3}. We have explicitly verified that graphs 2(b) and 2(c) vanish after momentum integration. The only nontrivial contributions come from graphs 2(a) and 2(d). The Feynman diagram 2(a) corresponds to the usual vacuum polarization of the photon observed in QED, with the additional VSR nonlocal corrections incorporated into the fermion propagator and vertex. Its analytical expression is given by
\begin{align}
S_{\text{eff}}^{(2,a)} & =\frac{i}{2}\mbox{Tr}\left[\frac{i}{i\cancel{\partial}+i\frac{m_{\psi}^{2}}{2}\frac{\cancel{n}}{n\cdot\partial}-m}ie\left(\cancel{A}-\frac{m_{\psi}^{2}}{2}\frac{\cancel{n}}{n\cdot\partial}(n\cdot A)\frac{1}{n\cdot\partial}\right)\right.\\
 & \times\left.\frac{i}{i\cancel{\partial}+i\frac{m_{\psi}^{2}}{2}\frac{\cancel{n}}{n\cdot\partial}-m}ie\left(\cancel{A}-\frac{m_{\psi}^{2}}{2}\frac{\cancel{n}}{n\cdot\partial}(n\cdot A)\frac{1}{n\cdot\partial}\right)\right],
\end{align}which written in momentum space result in
\begin{equation}
S_{\text{eff}}^{(2,a)}=\frac{i}{2}\int\frac{d^{4}q}{(2\pi)^{4}}\Pi_{\mu\nu}^{(2,a)}A^{\mu}(q)A^{\nu}(-q),
\end{equation}
where
\begin{equation}
\Pi_{(2,a)}^{\mu\nu}=-e^{2}\mbox{Tr}\int\frac{d^{4}p}{(2\pi)^{4}}\frac{i(\cancel{\tilde{p}}+m)}{p^{2}-\mu^{2}}\left(\gamma^{\mu}+\frac{m_{\psi}^{2}}{2}\frac{\cancel{n}}{n\cdot p}n^{\mu}\frac{1}{n\cdot u}\right)\frac{i(\cancel{\tilde{u}}+m)}{u^{2}-\mu^{2}}\left(\gamma^{\nu}+\frac{m_{\psi}^{2}}{2}\frac{\cancel{n}}{n\cdot u}n^{\nu}\frac{1}{n\cdot p}\right).
\end{equation}
%\begin{align}
%\Pi_{(2,a)}^{\mu\nu} & =\mbox{Tr}\int\frac{d^{4}p}{(2\pi)^{4}}\frac{i(\cancel{\tilde{p}}+m)}{p^{2}-\mu^{2}}\left(\gamma^{\mu}+\frac{m_{\psi}^{2}}{2}\frac{\cancel{n}}{n\cdot p}n^{\mu}\frac{1}{n\cdot u}\right)\\
% & \times\frac{i(\cancel{\tilde{u}}+m)}{u^{2}-\mu^{2}}\left(\gamma^{\nu}+\frac{m_{\psi}^{2}}{2}\frac{\cancel{n}}{n\cdot u}n^{\nu}\frac{1}{n\cdot p}\right).
%\end{align}
Following the procedure outlined earlier, we find
\begin{align}
\Pi_{\mu\nu}^{(2,a)} & =\frac{-ie^{2}}{12\pi^{2}(2-\omega)}\left[q_{\mu}q_{\nu}-q^{2}\left(\eta_{\mu\nu}+3m_{\psi}^{2}\frac{n_{\mu}n_{\nu}}{(n\cdot q)^{2}}\right)\right]\nonumber\\
 & +\frac{ie^{2}}{4\pi^{2}}\left\{ \frac{1}{3}\left(q_{\mu}q_{\nu}-q^{2}\eta_{\mu\nu}\right)\left[\gamma_{\text{E}}-\ln4\pi+6\mathcal{I}_{1}\right]\right.\nonumber\\
 & -\frac{m_{\psi}^{2}q^{2}}{n\cdot q}\left[n_{\mu}q_{\nu}+n_{\nu}q_{\mu}-(n\cdot q)\eta_{\mu\nu}\right]\mathcal{I}_{2}\nonumber\\
 & -\left.\frac{m_{\psi}^{2}q^{2}n_{\mu}n_{\nu}}{(n.q)^{2}}\left[\gamma_{\text{E}}-\ln4\pi+2q^{2}\mathcal{I}_{3}-\mathcal{I}_{4}\right]\right\},\label{Pi2a} 
\end{align}where we have defined the integrals:
\begin{equation}
\mathcal{I}_{1}\equiv\int_{0}^{1}dxx(1-x)\ln\left(\mu^{2}-q^{2}x(1-x)\right),
\end{equation}
\begin{equation}
\mathcal{I}_{2}\equiv\int_{0}^{1}dx\int_{0}^{1}dt\frac{x}{\mu^{2}-q^{2}x(1-x+tx)},
\end{equation}
\begin{equation}
\mathcal{I}_{3}\equiv\int_{0}^{1}dx\int_{0}^{1}dt\frac{x^{2}(t-2)(1-x+tx)}{\mu^{2}-q^{2}x(1-x+tx)},
\end{equation}
\begin{equation}
\mathcal{I}_{4}\equiv\int_{0}^{1}dx\int_{0}^{1}dt(1-6x+4tx)\ln\left(\mu^{2}-q^{2}x(1-x+tx)\right).
\end{equation}

Expression (\ref{Pi2a}), by itself, does not have a transverse structure as required by $U(1)$ gauge invariance of the photon field. To bring the photon self-energy into its desired form, we must include the full diagram contributions $\Pi_{\mu\nu}^{\text{(Total)}}=\Pi_{\mu\nu}^{(1,c)}+\Pi_{\mu\nu}^{(2,a)}$. Let us separate this sum into two parts. The one with the simplest structure is the divergent contribution, namely,
\begin{align}
\left.\Pi_{\mu\nu}^{\text{(Total)}}\right|_{\text{div}} & =\left.\Pi_{\mu\nu}^{(1,c)}\right|_{\text{div}}+\left.\Pi_{\mu\nu}^{(2,a)}\right|_{\text{div}}\\
 & =\frac{-ie^{2}}{12\pi^{2}(2-\omega)}\left(q_{\mu}q_{\nu}-q^{2}\eta_{\mu\nu}\right),
\end{align}which has the same form present in the usual QED. The second part is UV finite and exhibits a more intricate structure. To simplify the integrals $\mathcal{I}_{i}$, we expand the corresponding contributions as a power series in the external momenta before integrating over the parameters $x$ and $t$. The resulting expression can be written in the following form:
\begin{align}
\left.\Pi_{\mu\nu}^{\text{(Total)}}\right|_{\text{finite}} & =\left.\Pi_{\mu\nu}^{(1,c)}\right|_{\text{finite}}+\left.\Pi_{\mu\nu}^{(2,a)}\right|_{\text{finite}}\nonumber\\
 & =\frac{ie^{2}}{12\pi^{2}}\left\{ \left(\gamma_{\text{E}}+\ln(\frac{\mu^{2}}{4\pi})-\frac{1}{5}\frac{q^{2}}{\mu^{2}}-\frac{3}{140}\frac{q^{4}}{\mu^{4}}-\frac{1}{315}\frac{q^{6}}{\mu^{6}}-\cdots\right)\left(q_{\mu}q_{\nu}-q^{2}\eta_{\mu\nu}\right)\right.\nonumber\\
 & -\left.m_{\psi}^{2}\left(\frac{3}{2}\frac{q^{2}}{\mu^{2}}+\frac{5}{8}\frac{q^{4}}{\mu^{4}}+\frac{19}{60}\frac{q^{6}}{\mu^{6}}+\cdots\right)\left(\frac{n_{\mu}q_{\nu}+n_{\nu}q_{\mu}}{n\cdot q}-\frac{q^{2}n_{\mu}n_{\nu}}{(n\cdot q)^{2}}-\eta_{\mu\nu}\right)\right\}, 
\end{align}where $\cdots$ indicates terms of higher order in the power of $q^{2}/\mu^{2}$. In the end, the full one-loop vacuum polarization of the photon in the VSR-QED is manifestly transverse, as required by the Ward identity $q_{\mu}\Pi^{\mu\nu}(q)=0$. But only the UV finite part receives corrections due to the VSR-nonlocal terms. Note that a similar result was obtained recently in Ref. \cite{Haghgouyan:2022dvq} for the low-energy limit, $q^{2}\ll\mu^{2}$, in the context of Maxwell-Chern-Simons electrodynamics within the VSR framework.

The results obtained so far allow us to write the VSR effective Lagrangian for the gauge field $A_{\mu}$ as follows:
\begin{align}
\mathcal{L}_{\text{eff}}[A] & =\frac{1}{Z_{A}}\left(-\frac{1}{4}F_{\mu\nu}F^{\mu\nu}\right)+\frac{1}{2}m_{A}^{2}n_{\mu}\left(\frac{1}{n\cdot\partial}F^{\mu\nu}\right)n^{\alpha}\left(\frac{1}{n\cdot\partial}F_{\alpha\nu}\right)\nonumber \\
 & -\frac{e^{2}}{12\pi^{2}}\left\{ -\frac{1}{4}F_{\mu\nu}\left(\frac{1}{5}\frac{\Box}{\mu^{2}}-\frac{3}{140}\frac{\Box^{2}}{\mu^{4}}+\frac{1}{315}\frac{\Box^{3}}{\mu^{6}}-\cdots\right)F^{\mu\nu}\right.\nonumber \\
 & \left.-\frac{m_{\psi}^{2}}{2}n_{\mu}\left(\frac{1}{n\cdot\partial}F^{\mu\nu}\right)\left(\frac{3}{2}\frac{\Box}{\mu^{2}}-\frac{5}{8}\frac{\Box^{2}}{\mu^{4}}+\frac{19}{60}\frac{\Box^{3}}{\mu^{6}}+\cdots\right)n^{\alpha}\left(\frac{1}{n\cdot\partial}F_{\alpha\nu}\right)\right\},
\end{align}where
\begin{equation}
\frac{1}{Z_{A}}=1+\frac{e^{2}}{12\pi^{2}}C_{\text{div}},
\end{equation}with
\begin{equation}
C_{\text{div}}=\frac{1}{2-\omega}-\gamma_{\text{E}}-\ln\frac{\mu^{2}}{4\pi}.\label{Cdiv}
\end{equation}
In order to yield a divergence-free effective Lagrangian we perform a suitable change of scale, by defining the renormalized quantities:
\begin{equation}
A_{R}^{\mu}=Z_{A}^{-\frac{1}{2}}A^{\mu},
\end{equation}
\begin{equation}
e_{R}=Z_{A}^{\frac{1}{2}}e,
\end{equation}
\begin{equation}
m_{AR}=Z_{A}^{\frac{1}{2}}m_{A},
\end{equation}
Using the above scale transformations we get
\begin{align}
\mathcal{L}_{\text{eff}}[A] & =-\frac{1}{4}F_{R\mu\nu}F_{R}^{\mu\nu}+\frac{1}{2}m_{AR}^{2}n_{\mu}\left(\frac{1}{n\cdot\partial}F_{R}^{\mu\nu}\right)n^{\alpha}\left(\frac{1}{n\cdot\partial}F_{R\alpha\nu}\right)\nonumber \\
 & -\frac{e_{R}^{2}}{12\pi^{2}}\left\{ -\frac{1}{4}F_{R\mu\nu}\left(\frac{1}{5}\frac{\Box}{\mu^{2}}-\frac{3}{140}\frac{\Box^{2}}{\mu^{4}}+\frac{1}{315}\frac{\Box^{3}}{\mu^{6}}-\cdots\right)F_{R}^{\mu\nu}\right.\nonumber \\
 & \left.-\frac{m_{\psi}^{2}}{2}n_{\mu}\left(\frac{1}{n\cdot\partial}F_{R}^{\mu\nu}\right)\left(\frac{3}{2}\frac{\Box}{\mu^{2}}-\frac{5}{8}\frac{\Box^{2}}{\mu^{4}}+\frac{19}{60}\frac{\Box^{3}}{\mu^{6}}+\cdots\right)n^{\alpha}\left(\frac{1}{n\cdot\partial}F_{R\alpha\nu}\right)\right\} .
\end{align}
This effective Lagrangian retains the exact form of its classical counterpart, incorporating finite terms encompassing higher-derivative and VSR-nonlocal corrections. In particular, no additional counterterms were required to cancel the divergences, and the tensorial structure of these finite terms preserves both VSR and gauge symmetry.

Now we calculate the Kalb-Ramond self-energy from diagram 2(d) in Fig. \ref{Fig3}. Once again, expression (\ref{Seffn}) furnishes the analytical form of the corresponding effective action:
%\begin{equation}
%S_{\text{eff}}^{(2,d)}=\frac{i}{2}\mbox{Tr}\left[\frac{i}{i\cancel{\partial}+i\frac{m_{\psi}^{2}}{2}\frac{\cancel{n}}{n\cdot\partial}-m}i\frac{g}{12m}\sigma_{\mu\nu\lambda}\overset{\nsim}{H}^{\mu\nu\lambda}\frac{i}{i\cancel{\partial}+i\frac{m_{\psi}^{2}}{2}\frac{\cancel{n}}{n\cdot\partial}-m}i\frac{g}{12m}\sigma_{\alpha\beta\delta}\overset{\nsim}{H}^{\alpha\beta\delta}\right]
%\end{equation}
\begin{align}
S_{\text{eff}}^{(2,d)} & =\frac{i}{2}\mbox{Tr}\left[\frac{i}{i\cancel{\partial}+i\frac{m_{\psi}^{2}}{2}\frac{\cancel{n}}{n\cdot\partial}-m}i\frac{g}{12m}\sigma_{\mu\nu\lambda}\overset{\nsim}{H}^{\mu\nu\lambda}\right.\\
 & \times\left.\frac{i}{i\cancel{\partial}+i\frac{m_{\psi}^{2}}{2}\frac{\cancel{n}}{n\cdot\partial}-m}i\frac{g}{12m}\sigma_{\alpha\beta\delta}\overset{\nsim}{H}^{\alpha\beta\delta}\right].
\end{align}
Thus, we can write it in the momentum space as
\begin{equation}
S_{\text{eff}}^{(2,d)}=\frac{i}{2}\int\frac{d^{4}q}{(2\pi)^{4}}\Pi_{\mu\nu\lambda;\alpha\beta\delta}^{(2,d)}\overset{\nsim}{H}^{\mu\nu\lambda}(q)\overset{\nsim}{H}^{\alpha\beta\delta}(-q),
\end{equation}
where
\begin{equation}
\Pi_{\mu\nu\lambda;\alpha\beta\delta}^{(2,d)}=\left(\frac{ig}{12m}\right)^{2}\mbox{Tr}\int\frac{d^{4}p}{(2\pi)^{4}}\frac{i(\cancel{\tilde{p}}+m)}{p^{2}-\mu^{2}}\sigma_{\mu\nu\alpha}\frac{i(\cancel{\tilde{u}}+m)}{u^{2}-\mu^{2}}\sigma_{\alpha\beta\delta}.
\end{equation}
After making the Dirac traces and the momentum integration, we obtain
\begin{align}
\Pi_{(2,d)}^{\mu\nu\lambda;\alpha\beta\delta} & =\frac{ig^{2}\epsilon^{\mu\nu\lambda\sigma}\epsilon^{\alpha\beta\delta\rho}}{9m^{2}2^{2+2\omega}\pi^{\omega}}\Gamma(2-\omega)\left\{ 2q_{\sigma}q_{\rho}\int_{0}^{1}dx\frac{x(1-x)}{\left(\mu^{2}-q^{2}x(1-x)\right)^{2-\omega}}\right.\nonumber \\
 & -\frac{m_{\psi}^{2}q^{2}n_{\sigma}n_{\rho}}{(n\cdot q)^{2}}\int_{0}^{1}dx\int_{0}^{1}dt\frac{m_{\psi}^{2}x(2-\omega)+\mu^{2}-q^{2}x(1-x+tx)}{\left(\mu^{2}-q^{2}x(1-x+tx)\right)^{3-\omega}}\\
 & +\frac{m_{\psi}^{2}\left(n_{\sigma}q_{\rho}+n_{\rho}q_{\sigma}\right)}{n\cdot q}\int_{0}^{1}dx\int_{0}^{1}dt\frac{\mu^{2}+q^{2}x(1-x+tx)(1-\omega)}{\left(\mu^{2}-q^{2}x(1-x+tx)\right)^{3-\omega}}\nonumber \\
 & -\left.\eta_{\sigma\rho}\int_{0}^{1}dx\int_{0}^{1}dt\left[\frac{m_{\psi}^{2}q^{2}x(2-\omega)}{\left(\mu^{2}-q^{2}x(1-x+tx)\right)^{3-\omega}}-\frac{2\left(\mu^{2}-m_{\psi}^{2}-q^{2}x(1-x)\right)}{\left(\mu^{2}-q^{2}x(1-x)\right)^{2-\omega}}\right]\right\}. \nonumber 
\end{align}
As in the previous case, we can split the above result into divergent and finite parts as follows:
\begin{equation}
\left.\Pi_{(2,d)}^{\mu\nu\lambda;\alpha\beta\delta}\right|_{\text{div}}=\frac{ig^{2}\epsilon^{\mu\nu\lambda\sigma}\epsilon^{\alpha\beta\delta\rho}}{576\pi^{2}m^{2}(2-\omega)}\left[\frac{1}{3}\left(q_{\sigma}q_{\rho}+(6\mu^{2}-q^{2})\eta_{\sigma\rho}\right)+m_{\psi}^{2}\left(\frac{n_{\sigma}q_{\rho}+n_{\rho}q_{\sigma}}{n\cdot q}-\frac{q^{2}n_{\sigma}n_{\rho}}{(n\cdot q)^{2}}-2\eta_{\sigma\rho}\right)\right],
\end{equation}and
\begin{align}
&\left.\Pi_{(2,d)}^{\mu\nu\lambda;\alpha\beta\delta}\right|_{\text{finite}}=\nonumber\\
& =\frac{ig^{2}\epsilon^{\mu\nu\lambda\sigma}\epsilon^{\alpha\beta\delta\rho}}{576\pi^{2}m^{2}}\left\{ \frac{1}{3}q^{2}\eta_{\sigma\rho}-\frac{1}{3}(\gamma_{\text{E}}+\ln\frac{\mu^{2}}{4\pi})\left(q_{\sigma}q_{\rho}+(6\mu^{2}-q^{2})\eta_{\sigma\rho}\right)\right.\nonumber \\
 & -m_{\psi}^{2}\left[(\gamma_{\text{E}}+\ln\frac{\mu^{2}}{4\pi})\left(\frac{n_{\sigma}q_{\rho}+n_{\rho}q_{\sigma}}{n\cdot q}-\frac{q^{2}n_{\sigma}n_{\rho}}{(n\cdot q)^{2}}-2\eta_{\sigma\rho}\right)\right.\nonumber\\
 & \left.+\frac{q^{2}}{\mu^{2}}\left(\frac{q^{2}n_{\rho}n_{\sigma}}{3(n\cdot q)^{2}}-\frac{2\left(n_{\sigma}q_{\rho}+n_{\rho}q_{\sigma}\right)}{3n\cdot q}+\frac{5\eta_{\sigma\rho}}{6}\right)+\frac{q^{4}}{\mu^{4}}\left(\frac{3q^{2}n_{\rho}n_{\sigma}}{40(n\cdot q)^{2}}-\frac{9\left(n_{\sigma}q_{\rho}+n_{\rho}q_{\sigma}\right)}{40n\cdot q}+\frac{29\eta_{\sigma\rho}}{120}\right)+\cdots\right]\nonumber\\
 & \left.-m_{\psi}^{4}\left(\frac{1}{2}\frac{q^{2}}{\mu^{2}}+\frac{5}{24}\frac{q^{4}}{\mu^{4}}+\cdots\right)\frac{n_{\rho}n_{\sigma}}{(n\cdot q)^{2}}+\frac{q^{2}}{\mu^{2}}\left(\frac{q_{\rho}q_{\sigma}}{15}-\frac{q^{2}\eta_{\sigma\rho}}{30}\right)+\frac{q^{4}}{\mu^{4}}\left(\frac{q_{\rho}q_{\sigma}}{140}-\frac{q^{2}\eta_{\sigma\rho}}{420}\right)+\cdots\right\}.
\end{align}
It should be noted that both the divergent and finite parts receive radiative corrections due to VSR-nonlocal terms, in contrast to the pure Maxwell case. 

To determine the tensor structure of the quantum effective action and the necessary counterterms to renormalize it, we can utilize the identity
\begin{equation}
H_{\mu}^{\star}H_{\nu}^{\star}=\frac{1}{2}H_{\mu\alpha\beta}H_{\nu}^{\ \alpha\beta}-\frac{1}{6}\eta_{\mu\nu}H_{\alpha\beta\lambda}H^{\alpha\beta\lambda},
\end{equation}where $H_{\mu}^{\star}$ is the dual tensor corresponding to $H_{\mu\nu\lambda}$ and is given by
\begin{equation}
H_{\mu}^{\star}\equiv\frac{1}{3!}\epsilon_{\mu\alpha\beta\lambda}H^{\alpha\beta\lambda}.
\end{equation}
So we can write the effective Lagrangian density associated with the Kalb-Ramond field as 
\begin{align}
\mathcal{L}_{\text{eff}}[B] & =\frac{1}{Z_{B}}\left(-\frac{1}{12}\overset{\nsim}{H}_{\mu\nu\alpha}\overset{\nsim}{H}^{\mu\nu\alpha}\right)\nonumber\\
 & +\frac{g^{2}}{16\pi^{2}m^{2}}C_{\text{div}}\left\{ \frac{1}{12}\overset{\nsim}{H}_{\mu}^{\ \alpha\beta}(\partial^{\mu}\partial^{\nu})\overset{\nsim}{H}_{\nu\alpha\beta}\right.\nonumber\\
 & \left.+\frac{3}{12}m_{\psi}^{2}\left[\overset{\nsim}{H}_{\mu}^{\ \alpha\beta}\left(\frac{n^{\mu}n^{\nu}\Box}{(n\cdot\partial)^{2}}\right)\overset{\nsim}{H}_{\nu\alpha\beta}-\overset{\nsim}{H}_{\mu}^{\ \alpha\beta}\left(\frac{n^{\mu}\partial^{\nu}+n^{\nu}\partial^{\mu}}{n\cdot\partial}\right)\overset{\nsim}{H}_{\nu\alpha\beta}\right]\right\}\nonumber \\
 & +\text{finite terms},\label{LeffB}
\end{align}where
\begin{equation}
\frac{1}{Z_{B}}=1-\frac{g^{2}\mu^{2}}{8\pi^{2}m^{2}}C_{\text{div}},
\end{equation}with $C_{\text{div}}$ defined in Eq. (\ref{Cdiv}). 

We note that from the second and third terms in Eq. (\ref{LeffB}), it is necessary to introduce new counterterms into the classical action in order to renormalize it within the minimal subtraction scheme. This is an expected result since it is well-known that nonminimal couplings can lead to higher-derivative divergences in the quantum effective action. Hence, at this point, we may interpret our model as an effective theory that is valid in the low-momentum limit, where $q^2 \ll m^{2}$. In this regime, we can make the approximations $\mu^{2}\approx m^{2}$ with $m_{\psi}^{2}\ll1$. Thus we have
\begin{equation}
\frac{1}{Z_{B}}\approx1-\frac{g^{2}}{8\pi^{2}}C_{\text{div}}
\end{equation}
and by defining a renormalized field $B_{R}^{\mu\nu}=Z_{B}^{-\frac{1}{2}}B^{\mu\nu}$ and a renormalized coupling constant $g_{R}=Z_{B}^{\frac{1}{2}}g$, we get
\begin{equation}
\mathcal{L}_{\text{eff}}[B] \approx-\frac{1}{12}\overset{\nsim}{H}_{R\mu\nu\alpha}\overset{\nsim}{H}_{R}^{\mu\nu\alpha}+\text{finite terms},
\end{equation}
which has same form as the classical free Kalb-Ramond Lagrangian in the VSR context. Moreover, in the limit of $m_{B} \to 0$, we recover the Lorentz covariant result obtained in Ref. \cite{HariDass:2001dp} at leading order in $H^{2}$, with the same values for the renormalized quantities.

%%%%%%%%%%%%%%%%%%%%%%%%%%%%%%%%%%%%%%%%%%%%%%%%%%%%%%%%%%%%%%%%%%%%%%
\section{Conclusions\label{sec5}}

In this work, we propose an extension of Maxwell and Kalb-Ramond electrodynamics in the presence of fermionic matter fields in a $SIM(2)$-gauge invariant manner, incorporating both minimal and nonminimal couplings. In the free case, the VSR-Kalb-Ramond field is equivalent to a single massive real scalar field with one polarization. Moreover, we have calculated the VSR-effective action for the Maxwell and Kalb-Ramond field strengths using the Alfaro-Mandelstam-Leibbrandt prescription \cite{Alfaro:2016pjw}, which accounts for UV/IR mixing divergences in the one-loop Feynman integrals. The induced quantum corrections include higher-derivative terms that preserve the VSR-nonlocal tensor structure. Specifically, the finite terms in the Maxwell sector exhibit the VSR-nonlocal tensor structure, while the divergent part retains Lorentz covariance. On the other hand, in the Kalb-Ramond sector, both finite and divergent terms incorporate VSR-nonlocal corrections. In the latter case, the counterterms necessary to cancel the divergences are absent in the bare Lagrangian, and additional counterterms must be introduced into the classical Lagrangian to absorb these divergences and render the theory well-defined. However, a renormalized effective action can be achieved in the weak-energy limit, similar to what has been obtained in the literature for the Lorentz-covariant case of constant $F$ and $H$ field strengths \cite{HariDass:2001dp}.

%%%%%%%%%%%%%%%%%%%%%%%%%%%%%%%%%%%%%%%%%%%%%%%%%%%%%%%%%%%%%%%%%%%%%%%%%%
\section*{Acknowledgments}
\hspace{0.5cm} The authors thank the Funda\c{c}\~{a}o Cearense de Apoio ao Desenvolvimento Cient\'{i}fico e Tecnol\'{o}gico (FUNCAP), the Coordena\c{c}\~{a}o de Aperfei\c{c}oamento de Pessoal de N\'{i}vel Superior (CAPES), and the Conselho Nacional de Desenvolvimento Cient\'{i}fico e Tecnol\'{o}gico (CNPq), Grant no. 200879/2022-7 (RVM). 
This work is also supported by the Spanish Agencia Estatal de  Investigaci\'on (grant PID2020-116567GB-C21 funded by MCIN/AEI/10.13039/501100011033 and ERDF A way of making Europe) and by the project PROMETEO/2020/079 (Generalitat Valenciana). R. V. Maluf thanks the Departament of Theoretical Physics \& IFIC  of the University of Valencia - CSIC for the kind hospitality.

%%%%%%%%%%%%%%%%%%%%%%%%%%%%%%%%%%%%%%%%%%%%%%%%%%%%%%%%%%%%%%%%%%%%%%%%%%

\end{document}